\documentclass[useAMS,usenatbib] {mn2e}

\usepackage{times}
\usepackage{graphics,epsfig}
\usepackage{graphicx}
\usepackage{amsmath}
\usepackage{amssymb}
\usepackage{comment}
\usepackage{supertabular}
\usepackage{lscape}

\newcommand{\mb}  {\ensuremath{\rm M_{B}}}

\newcommand{\nh}  {\ensuremath{\rm N_{HI}}}

\newcommand{\kms}{km~s$^{-1}$}

\newcommand{\ha}{\ensuremath{{\rm H}\alpha} }

\newcommand{\ms}{\ensuremath{\rm M_{\odot}}}

\newcommand{\acc}{\ensuremath{\rm atoms~cm^{-2}}}

\newcommand{\HI}{H{\sc i }}
\newcommand{\hi}{H{\sc i }}

\title {FIGGS2: An \HI survey of extremely faint irregular galaxies}
\author [Patra et al.]{
Narendra Nath Patra$^{1}$ \thanks {E-mail: narendra@ncra.tifr.res.in},
	Jayaram N. Chengalur$^{1}$ \thanks {E-mail: chengalu@ncra.tifr.res.in},
	Igor D. Karachentsev$^{2}$ \thanks {E-mail: ikar@sao.ru},
	\and Margarita E. Sharina$^2$ \\
	$^{1}$ NCRA-TIFR, Post Bag 3, Ganeshkhind, Pune 411 007, India \\ 
	$^{2}$ Special Astrophysical Observatory, Nizhnij Arkhyz, Karachai-Cherkessia 369167, Russia
}
\date {}
\begin {document}
\maketitle
\label{firstpage}

\begin{abstract}

We present the observations and first results from the FIGGS2 survey.
FIGGS2 is an extension of the earlier Faint Irregular Galaxies GMRT survey (FIGGS) towards faint luminosity end. The sample consists of 20 galaxies of which 15 were detected in \HI 21cm line using the Giant Meter-wave Radio Telescope (GMRT). The median blue band magnitude of our sample is $\sim -11.6$, which is more than one magnitude fainter than earlier FIGGS survey. From our GMRT observations we find that, for many of our sample galaxies, the \HI disks are offset from their optical disks. The \HI diameters of the FIGGS2 galaxies show a tight correlation with their \HI mass. The slope of the correlation is 2.08$\pm$0.20 similar to what is found for FIGGS galaxies. We also find that for almost all galaxies, the \HI disks are larger than the optical disks which is a  common trend for dwarf or spiral galaxies. The mean value of the ratio of \HI to optical diameter is $\sim$ 1.54.


\end{abstract}

\section{Introduction}

There are a number of issues that make systematic studies of dIrr galaxies particularly interesting. Firstly, in hierarchical models of galaxy formation, small objects form first and merge together to form larger objects. In that sense, nearby dwarf galaxies are the closest analogues to the galaxies in the early universe. Secondly, the ISM of dwarf galaxies have low metallicity. In this sense too, they are analogous to high redshift galaxies, and serve as a nearby laboratory for the study of gas and star formation in environments with low dust and low metallicity \citep{roychowdhury09, roychowdhury11}. This is in part responsible for the increasing number of recent surveys of dwarf galaxies, e.g. FIGGS \citep{begum08c}, SHIELD \citep{cannon11}, VLA-ANGST \citep{ott12}, LITTLE-THINGS \citep{hunter12}. 
 
In this paper we describe an extension to the FIGGS \citep{begum08c} survey. This extension focuses on galaxies with fainter luminosities and smaller \hi\ masses. We present here the results of our \HI observations of 20 very faint galaxies with the Giant Meterwave Radio Telescope (GMRT). In \S 2 we describe our sample, in \S 3 we explain the main science drivers of the survey, in \S 4 we present the observations and data analysis and finally in \S 5 we present the results and discussion.

\section{Sample}

The FIGGS2 survey was planned as an extension of the Faint Irregular Galaxy GMRT Survey (FIGGS) \citep{begum08c}. The FIGGS sample was based on the 2004 version of a compilation of nearby galaxies (Catalogue of Nearby Galaxies \citep{karachentsev04}). Since then there has been an almost two fold increase in the known number of faint galaxies in the local neighbourhood, thanks to surveys like the SDSS \citep{abazajian09} and ALFALFA \citep{giovanelli05a}. In the FIGGS2 survey we focus on the faintest end of the galaxy spectrum, viz. galaxies with $\mb \gtrsim -12$. The cutoff magnitude for the FIGGS sample was $\mb = -14.5$ and the sample contained $\sim 11$ galaxies fainter than $\mb = -12$. The FIGGS2 sample consists of 20 galaxies with $\rm M_B \gtrsim -12.0$, and $S_{\rm HI} \gtrsim 0.5$ Jy~km/s which combined with FIGGS galaxies leads to a $\sim 3$ times larger sample of galaxies fainter than $\rm M_B = -12.0$ than was earlier available. We note that revision to the distance and other observable parameters have resulted in two of our galaxies now having \mb~slightly larger than -12. The galaxies were selected from the Updated Nearby Galaxy Catalog (UNGC) \citep{karachentsev04} as per the telescope scheduling constraints. FIGGS sample consists of 66 galaxies out of $\sim $ 230 galaxies in the UNGC catalog satisfying selection criteria of FIGGS, whereas, 15 galaxies were observed with the GMRT as part of FIGGS2, out of $\sim$ 85 galaxies in NGC catalogue which satisfy selection criteria of FIGGS2. Most of the remaining objects (unobserved within FIGGS$+$FIGGS2) reside on the southern sky bellow the GMRT horizon.

 In Table~\ref{table1_figgs} we list a few general properties of our sample galaxies. The columns are as follows: column (1): Galaxy name, column (2) and (3): the equatorial coordinates (J2000), column (4): Distance in Mpc, column (5): the methods used to determine the distances to the galaxies, - by the tip of the red giant branch (TRGB), by the Hubble velocity-distance relation ($H_0$ = 73 km/s/Mpc) (h), from galaxy membership (mem), column (6) the absolute blue magnitude (extinction corrected), column (7): log of \HI mass, column (8) Heliocentric radial velocity, column (9): The Holmberg diameter, column (10): inclination derived from optical photometry (assuming an intrinsic thickness of 0.42 \citep{roychowdhury13}). The data presented in Table~\ref{table1_figgs} were taken from \citep{karachentsev13,karachentsev01,makarov03,huchtmeier00,huchtmeier09}. The first 15 galaxies in Table~\ref{table1_figgs} were detected in our GMRT observations, whereas the last five galaxies (separated by an empty line) were not detected.


\begin{table*}
\caption{Sample galaxy properties}
\begin{tabular}{|l|c|c|c|c|c|c|c|c|c|c|}
\hline
Galaxy & $\alpha$ (J2000) & $\delta$ (J2000) & Distance & Method & $\rm M_B$ & $\rm \log M_{HI}$ & $\rm V_{hel}$ & $\rm D_{opt}$ & $\rm i_{opt}$ \\
       & (hhmmsss) & ($^o~^{\prime}~^{\prime \prime}$) & (Mpc) & & (mag) & ($\rm M_{\odot}$) & \kms & (arcmin) & ($^o$) \\
(1) & (2) & (3) & (4) & (5) & (6) & (7) & (8) & (9) & (10) \\      
       
\hline
AGC112521 & $014107.9$ & $+271926$ & $6.08$ & $TRGB$ & $-11.4$ & $6.75$ & 274 & 0.60 & $67$\\ 
KK15 & $014641.6$ & $+264805$ & $8.67$ & $TRGB$ & $-11.8$ & $7.21$ & 366 & 0.59 & $90$\\ 
KKH37 & $064745.8$ & $+800726$ & $3.44$ & $TRGB$ & $-11.6$ & $6.71$ & 11 & 1.15 & $55$\\ 
KKH46 & $090836.6$ & $+051732$ & $6.70$ & $TF$ & $-12.3$ & $7.44$ & 598 & 0.60 & $34$\\ 
UGC04879 & $091602.2$ & $+525024$ & $1.36$ & $TRGB$ & $-11.9$ & $5.98$ & -25 & 3.24 & $66$\\ 
LeG06 & $103955.7$ & $+135428$ & $10.40$ & $mem$ & $-11.9$ & $6.85$ & 1007 & 0.63 & $57$\\ 
KDG073 & $105257.1$ & $+693245$ & $3.91$ & $TRGB$ & $-10.9$ & $6.56$ & 116 & 1.20 & $38$\\ 
VCC0381 & $121954.1$ & $+063957$ & $4.71$ & $h$ & $-11.7$ & $7.14$ & 480 & 0.78 & $26$\\ 
KK141 & $122252.7$ & $+334943$ & $7.78$ & $h$ & $-11.5$ & $7.20$ & 573 & 0.40 & $45$\\ 
KK152 & $123324.9$ & $+332105$ & $6.90$ & $TF$ & $-13.0$ & $7.54$ & 838 & 1.07 & $83$\\ 
UGCA292 & $123840.0$ & $+324600$ & $3.85$ & $TRGB$ & $-11.9$ & $7.49$ & 308 & 1.02 & $52$\\ 
BTS146 & $124002.1$ & $+380002$ & $8.50$ & $TF$ & $-12.2$ & $6.97$ & 446 & 0.34 & $67$\\ 
LVJ1243+4127 & $124355.7$ & $+412725$ & $6.09$ & $h$ & $-11.8$ & $7.02$ & 402 & 1.38 & $83$\\ 
KK160 & $124357.4$ & $+433941$ & $4.33$ & $TRGB$ & $-10.9$ & $6.60$ & 293 & 0.59 & $47$\\ 
KKH86 & $135433.6$ & $+041435$ & $2.61$ & $TRGB$ & $-10.3$ & $5.92$ & 287 & 0.85 & $51$\\ 
\\
LeG18 & $104653.3$ & $+124440$ & $10.40$ & $mem$ & $-11.3$ & $7.15$ & 636 & 0.28 & $47$\\ 
KDG90 & $121457.9$ & $+361308$ & $2.98$ & $TRGB$ & $-11.6$ & $7.66$ & 280 & 1.55 & $33$\\ 
LVJ1217+4703 & $121710.1$ & $+470349$ & $7.80$ & $mem$ & $-11.0$ & $7.38$ & 394 & 0.30 & $46$\\ 
KK138 & $122158.4$ & $+281434$ & $6.30$ & $mem$ & $-10.4$ & $6.81$ & 449 & 0.42 & $64$\\ 
KK191 & $131339.7$ & $+420239$ & $8.28$ & $TRGB$ & $-11.4$ & $7.59$ & 371 & 0.42 & $18$\\
\hline
\end{tabular}
\label{table1_figgs}
\end{table*}

\begin{figure*}
\begin{center}
\begin{tabular}{cc}
\resizebox{70mm}{!}{\includegraphics{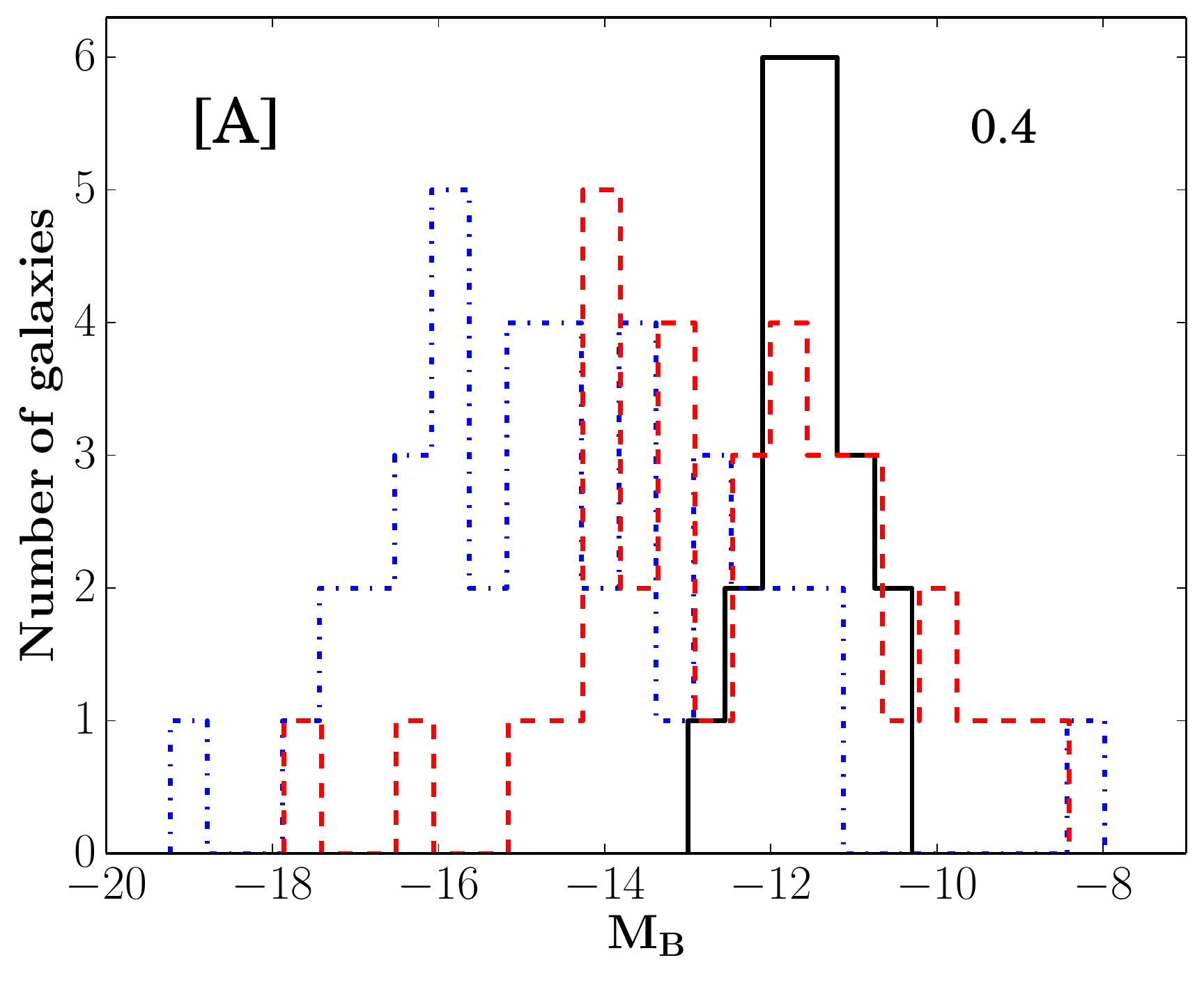}} &
\resizebox{72mm}{!}{\includegraphics{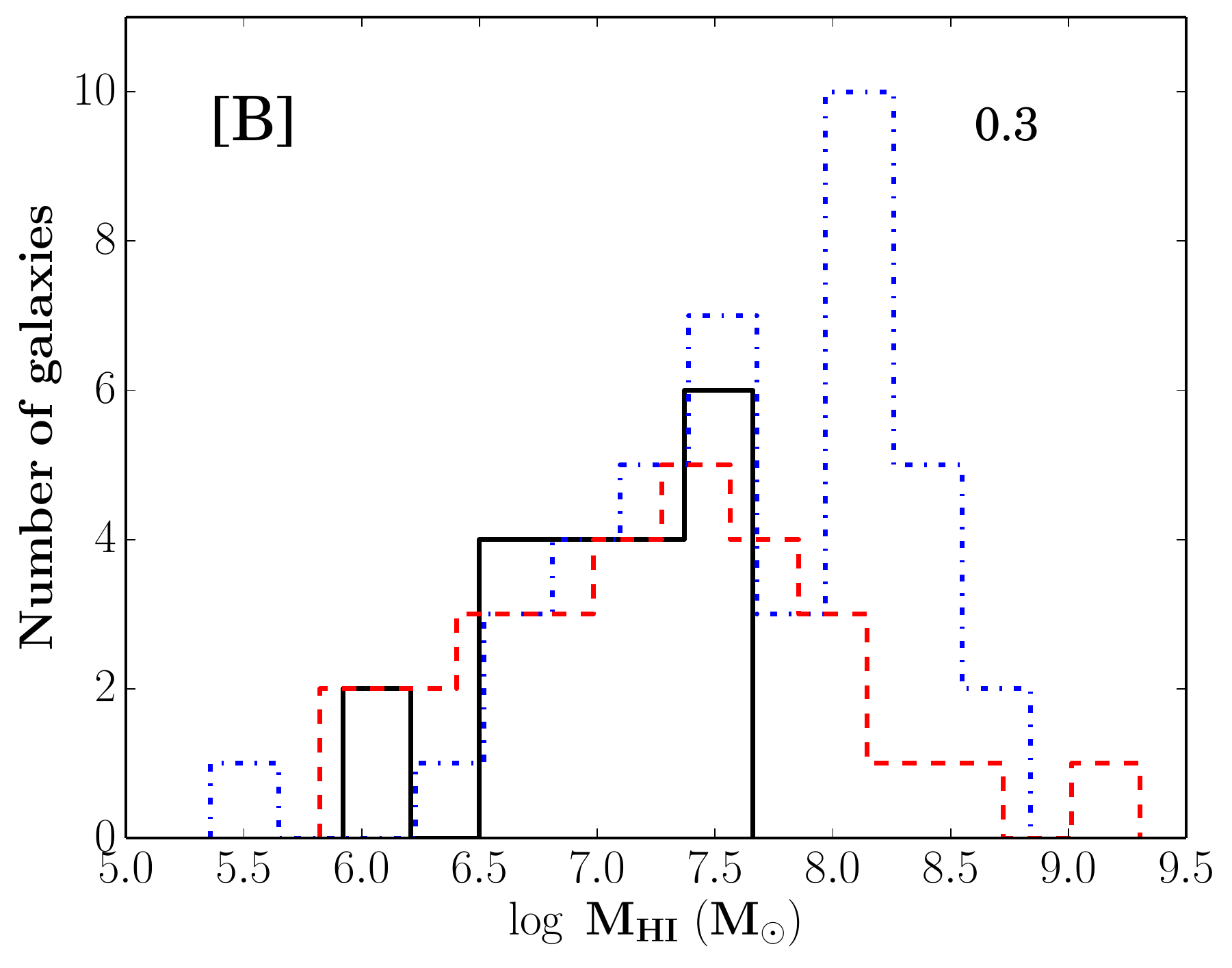}} \\
\resizebox{70mm}{!}{\includegraphics{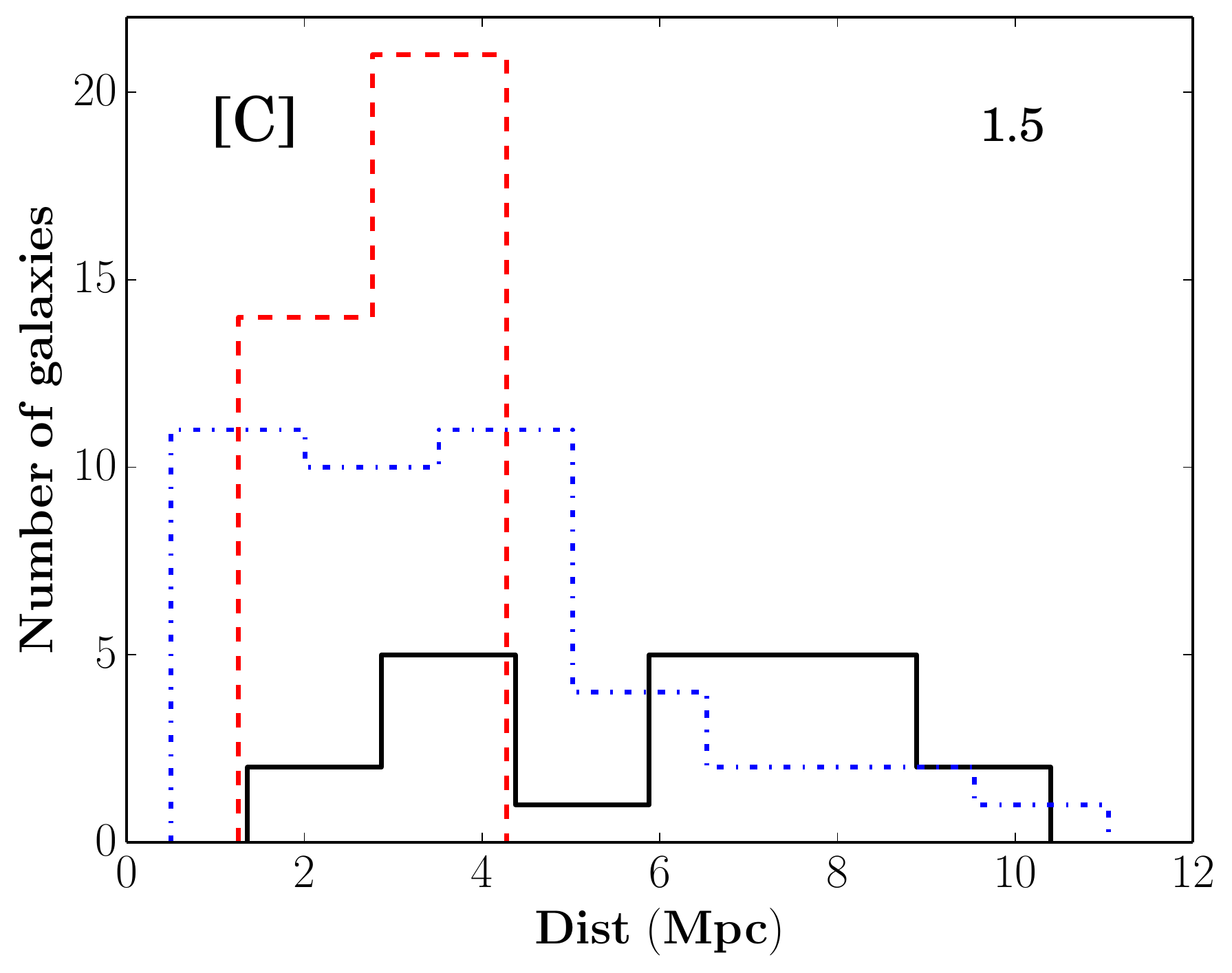}} &
\resizebox{70mm}{!}{\includegraphics{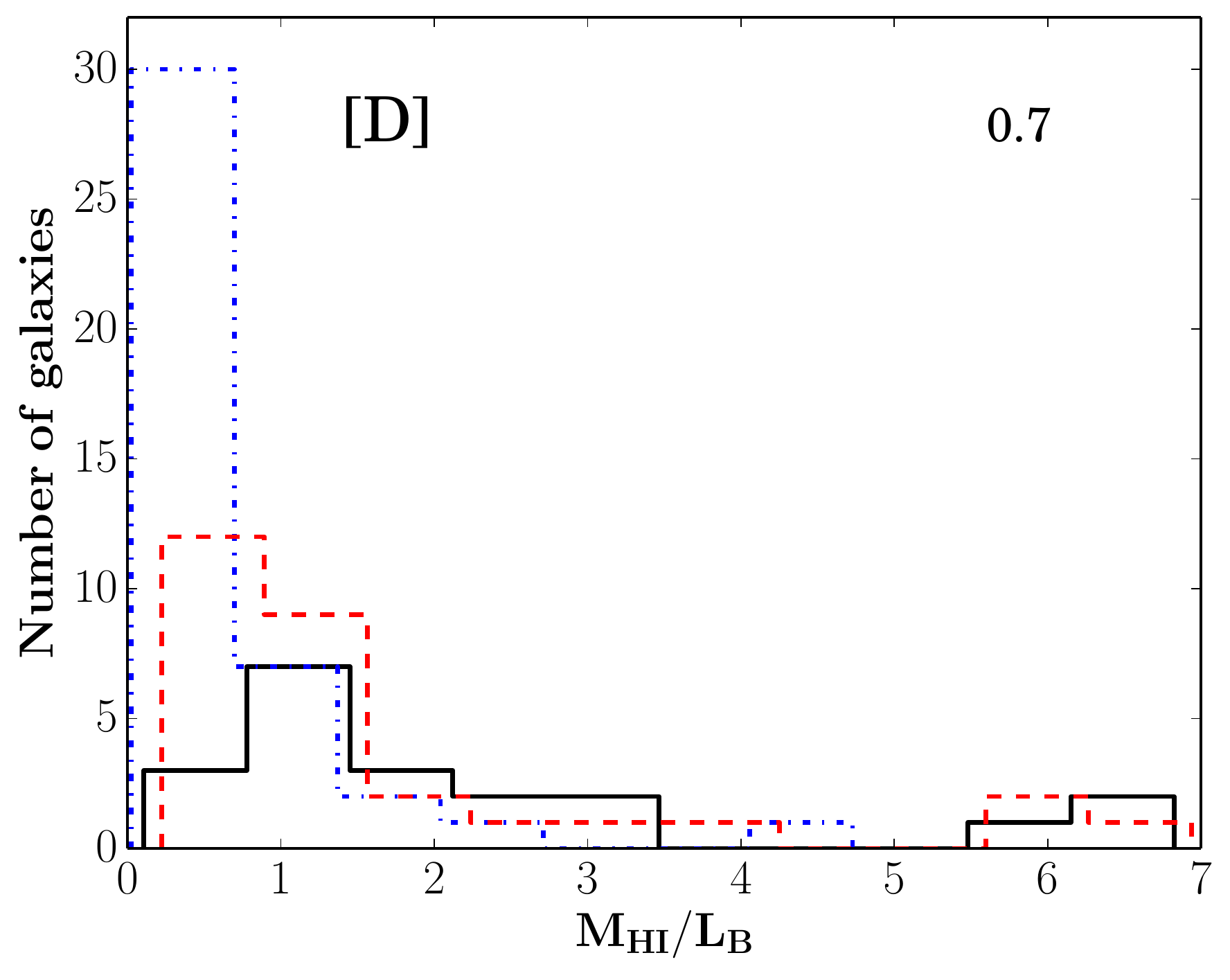}} \\
\end{tabular}
\end{center}
\caption{Histograms of different global properties of our sample galaxies. The solid black histograms represent FIGGS2 data (taken form Tab.~\ref{table1_figgs}). For comparison we also plot data from two major surveys of dwarf galaxies, namely, the VLA-ANGST and the LITTLE-THINGS survey. The blue dashed-dotted histograms represent data from the LITTLE-THINGS survey whereas the red dashed histograms are for VLA-ANGST survey. For consistency we have kept the bin width of the histograms same for all three surveys. The bin widths are quoted at the top right corners of the respective panels. Panel [A] shows the histograms of extinction corrected absolute blue magnitude ($\rm M_B$), panel [B] represents the histograms of $\log$ of \HI mass, panel [C] shows the histograms of distances and in panel [D] we show the histograms of the \HI mass to blue luminosity ratio ($\rm M_{HI}/L_B$). }
\label{histograms}
\end{figure*}

\begin{figure*}
\begin{center}
\begin{tabular}{ccc}
\resizebox{180mm}{!}{\includegraphics{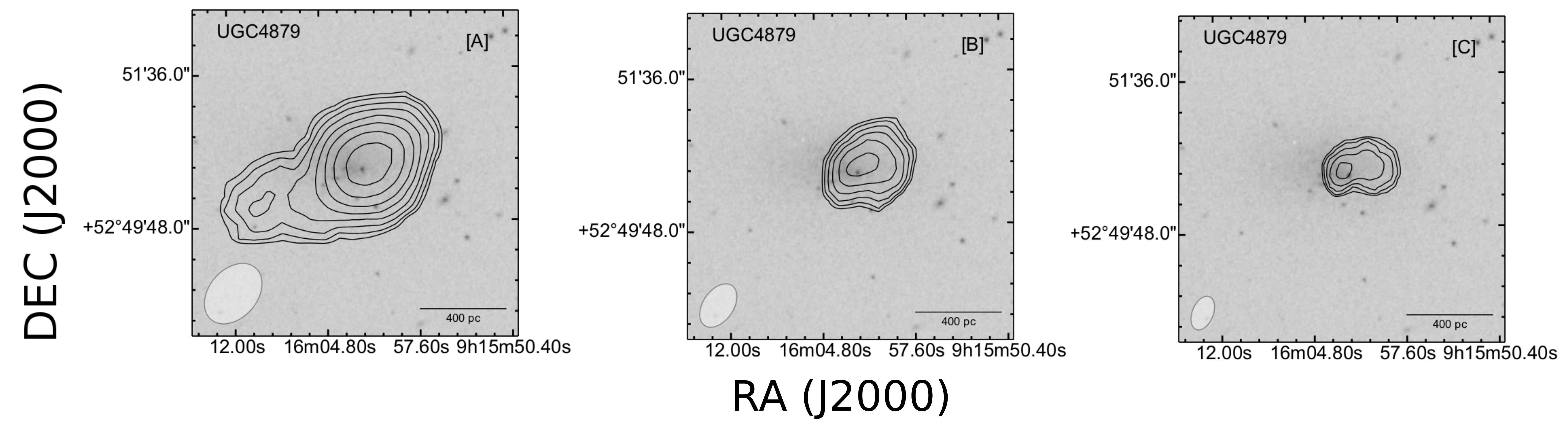}}
\end{tabular}
\end{center}
\caption{Integrated \HI emission from  UGC~4879 (contours) at different spatial resolutions overlayed on a DSS B band image (greyscales). The resolution of the images are  $48^{''}\times34^{''}$ (panel [A]), $34^{''}\times 21^{''}$ (panel [B]), $25^{''}\times14^{''}$ (panel [C]). The contour levels are (1, 1.4, 2, 2.8, ..) 2$\rm \times 10^{19}$ in panel [A], (1, 1.4, 2, 2.8, ..) 6$\rm \times 10^{19}$ in panel [B] and (1, 1.4, 2, 2.8, ..) 1$\rm \times 10^{20}$ in panel [C].}
\label{momres}
\end{figure*}

In Figure~\ref{histograms} we plot histograms of various global properties of our sample galaxies. To compare our survey with other major surveys, we plot histograms of sample galaxy properties of two major surveys of dwarf galaxies namely, the LITTLE-THINGS survey \citep{hunter12} and the VLA-ANGST survey \citep{ott12}. The solid black histograms in Fig.~\ref{histograms} represents FIGGS2 survey data, whereas the blue dashed-dotted and the red dashed histograms represent the LITTLE-THINGS and the VLA-ANGST data respectively. In panel [A] we plot the histograms of absolute blue magnitude $\rm M_B$, panel [B] shows the histograms of $\log$ of \HI mass, panel [C] and [D] shows the histograms of distances to the sample galaxies and the \HI mass to blue luminosity ratio ($\rm M_{HI}/L_B$) respectively. Since the distances to some of our galaxies have been updated after the sample selection was done, the
estimated luminosities of some of our sample galaxies are brighter than the sample selection cut-off of $\rm M_B \gtrsim -12.0$. Nonetheless,  the  median $\rm M_B$ of the sample is $\rm -11.6$, which is more than one magnitude fainter than the median of the FIGGS sample. Panel [B] (solid black line) shows the histogram of $\log$ of \HI mass of our sample galaxies. The median \HI mass of our sample galaxies is $\sim 8 \times 10^6$ \ms\ which is also about an order of magnitude lower than the median \HI mass of FIGGS sample. From Figure~\ref{histograms} one can see that our sample spans $\sim$ 3 magnitude in brightness (a factor of $\sim$ 12) and $\sim$ 2 orders of magnitude in \HI mass. We also note that our sample galaxies are concentrated around the low luminosity tail of the LITTLE-THINGS or the VLA-ANGST survey.

\section{Science drivers for FIGGS2}

The primary goal of the FIGGS2 survey was to extend the previous FIGGS survey towards the fainter end and enrich the multi wave length data base to address several science questions. A few of the science drivers of FIGGS2 are discussed below.

Much of what we know about gas in the high redshift universe comes from the study of absorption line systems seen in front of bright quasars, i.e. the so called Damped Lyman-$\alpha$ absorption systems (DLAs). Although such studies allow one to inventory the total amount of atomic gas as a function of redshift, because the information received is limited to that along the pencil beam illuminated by the quasar, the nature of the host population of these systems remains unclear. An interesting question is as to whether their properties resemble that of the local dwarf galaxy population. One quantitative way of checking this is is to use data from surveys like the FIGGS and FIGGS2 surveys to see whether the column density distribution function of DLAs matches that observed in local dwarf galaxies \citep[e.g.][]{patra13}.  

The neutral ISM and its connection with the star-formation in gas-rich dwarf irregular galaxies has been a major area of interest for a long time. Star formation in these low dust, low metallicity environments is expected to proceed differently than in spiral galaxies. Though a number of studies using FIGGS data have already explored many aspects of star formation \citep[see for example,][]{roychowdhury09, roychowdhury11}, yet a number of interesting questions still remain to be answered; like star formation feedback and its effect on star-formation in smallest scales, abundance of the different ISM phases and its connection with star formation etc. Very often the total measured \ha\ emission in these galaxies can be accounted for by only a few massive stars. Due to very shallow potential well of these galaxies, the ISM and cold gas are expected to be strongly affected by star formation feedback. A comparison of the \HI and optical morphologies could allow one to examine the consequences of this feedback in the smallest gas-rich galaxies.

Another area of interest is in the phase structure of the atomic gas in these galaxies. In our own galaxy the atomic ISM is believed to have two stable phases that co-exist in pressure equilibrium, i.e. a dense cold phase (the Cold Neutral Medium) and a warm diffuse phase (the Warm Neutral Medium). There is also increasing evidence that a significant fraction of atomic gas is a phase with intermediate temperature, which would be thermally unstable. There have been several studies aimed at trying to understand the phase structure of the atomic gas in dwarf galaxies, and one would like to extend such studies to the smallest star forming units known. A related question would be as to what the connections, if any, are between the CNM phase and star formation in dwarf galaxies (e.g. \citep{patra16}).

Another area of interest is to the structure of the dark matter halo and its influence on the structure and dynamics of dwarf galaxies \citep{banerjee08,banerjee10,sahakanak14}. The vertical structure and scale-height of galaxies is determined by the hydrostatic equilibrium between different galactic components \citep[e.g.][]{narayan02a} embedded in the dark matter halo. This vertical hydrostatic equilibrium decides in turn the thickness and the vertical structure of the galactic disk. Observationally it is found that the gas disks of small gas-rich galaxies (like our sample) are thicker than normal spirals \citep{roychowdhury10}. However a complete theoretical understanding of this higher thickness and the vertical structure of the gas disc of dwarf galaxies is not yet available. Similarly the presence of non-axi-sysmmetric structures also has implications for the dark matter distribution \citep{banerjee13}. One of the aims of this survey is to provide data for studies vertical structure of gas disks, which in turn can be used to constrain the distribution of the dark matter and the gas velocity dispersion \citep{patra14}.








\begin{figure*}
\begin{center}
\begin{tabular}{ccc}
\resizebox{180mm}{!}{\includegraphics{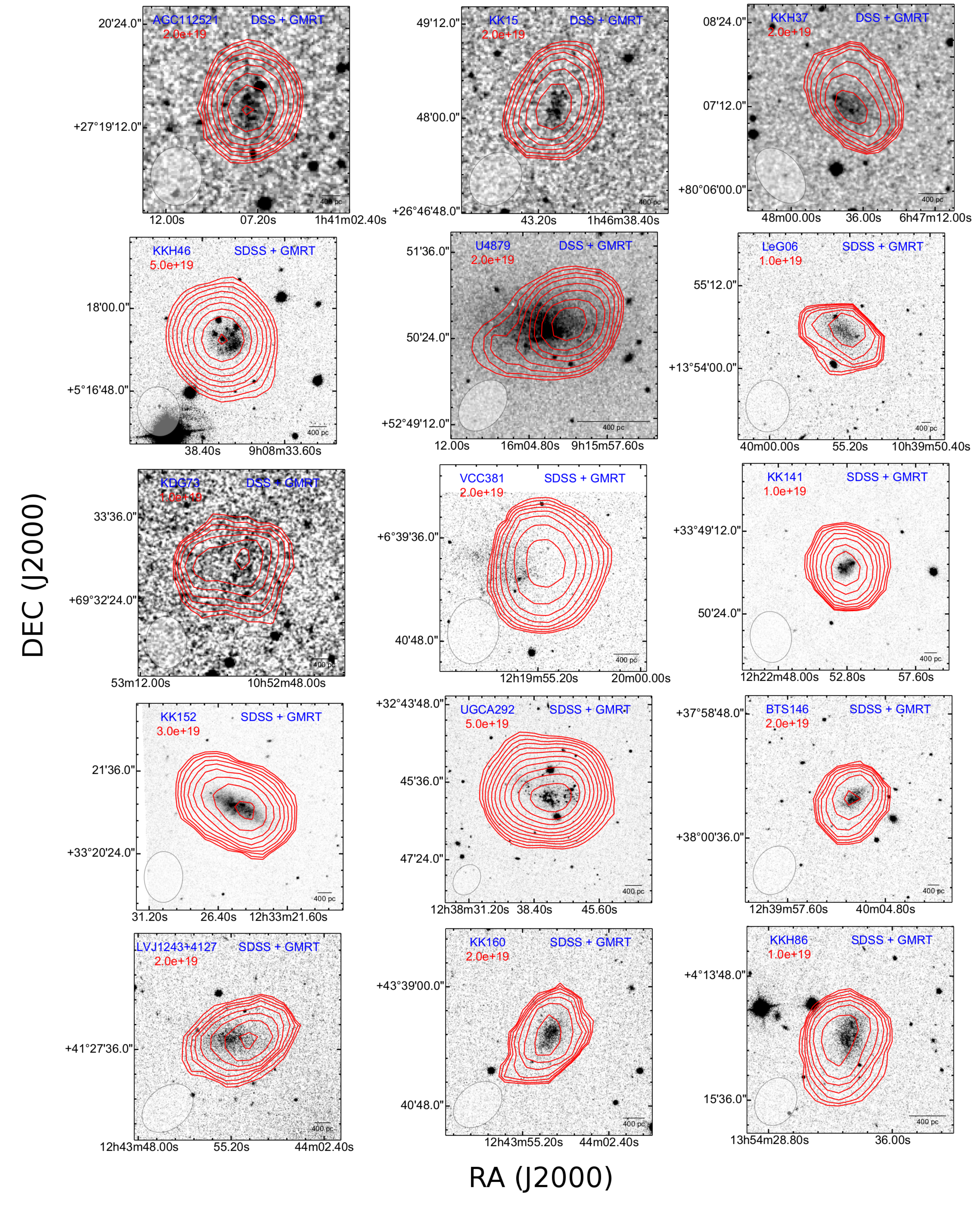}}
\end{tabular}
\end{center}
\caption{Overlays of the integrated \HI emission (contours) on the optical image for the FIGGS2 galaxies. The optical images were taken from SDSS \citep{abazajian09} (`g' filter; $\lambda$ centred at $\sim$ 4770$\rm \AA$) if available, else DSS images (red filter $\lambda$ at $\sim$ 6450 $\rm \AA$) were used.  The lowest \HI contour levels are quoted on the top left of the respective panels in the units of \acc. The successive contours are separated by a factor of $\sqrt 2$.} 
\label{ovrplot}
\end{figure*}

\begin{figure*}
\begin{center}
\begin{tabular}{ccc}
\resizebox{180mm}{!}{\includegraphics{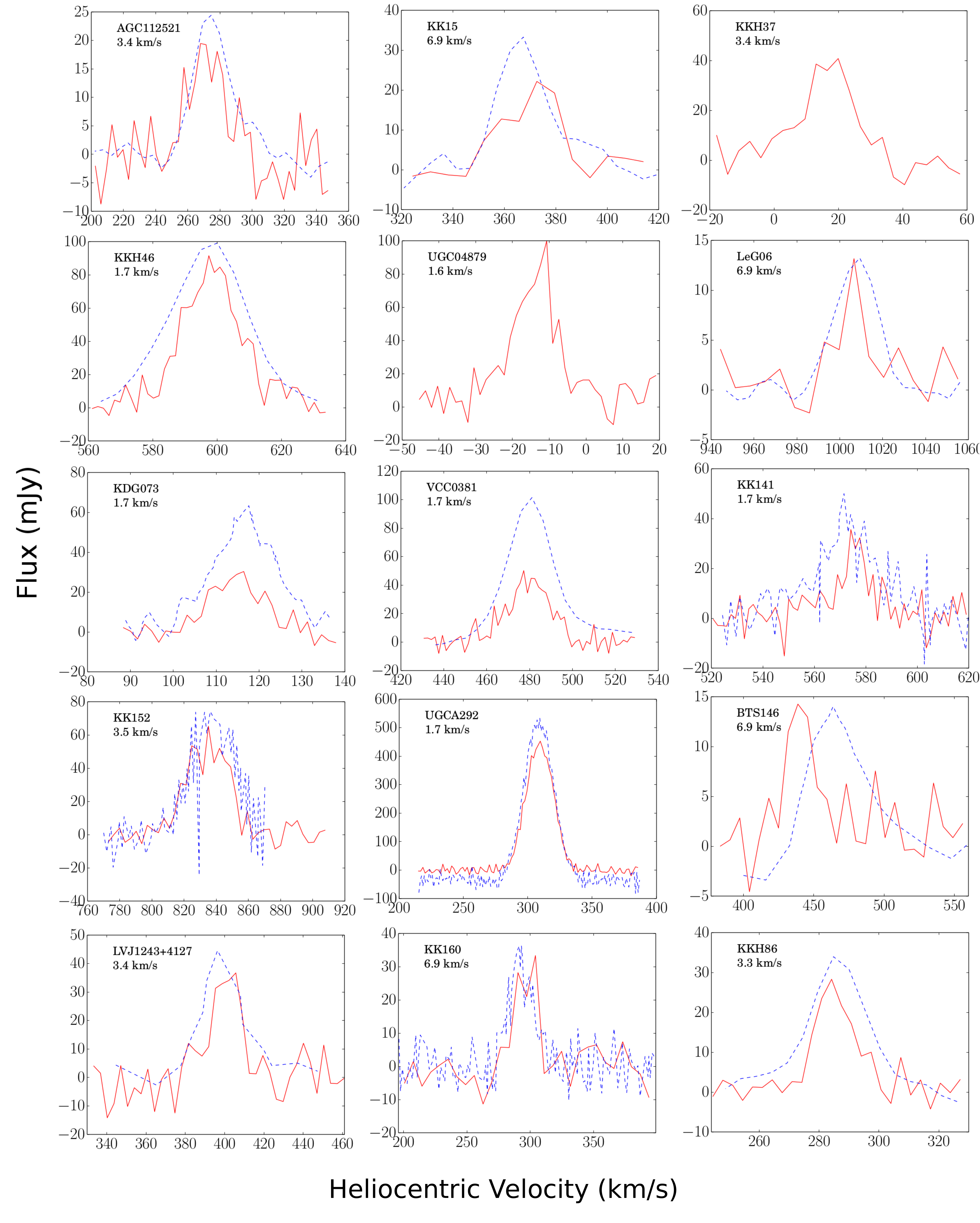}}
\end{tabular}
\end{center}
\caption{Global \HI spectra of our sample galaxies (red solid line) plotted along with the single-dish spectra (blue dashed line). To increase the SNR, multiple channels were collapsed together wherever necessary. The velocity resolution used is quoted in the respective panels. We note that in most of the cases, GMRT observation recovers less flux as compared to single-dish flux. The single-dish spectra for KKH37 and UGC04879 is not available in literature.}
\label{spec}
\end{figure*}

\begin{figure*}
\begin{center}
\begin{tabular}{cccc}
\resizebox{190mm}{!}{\includegraphics{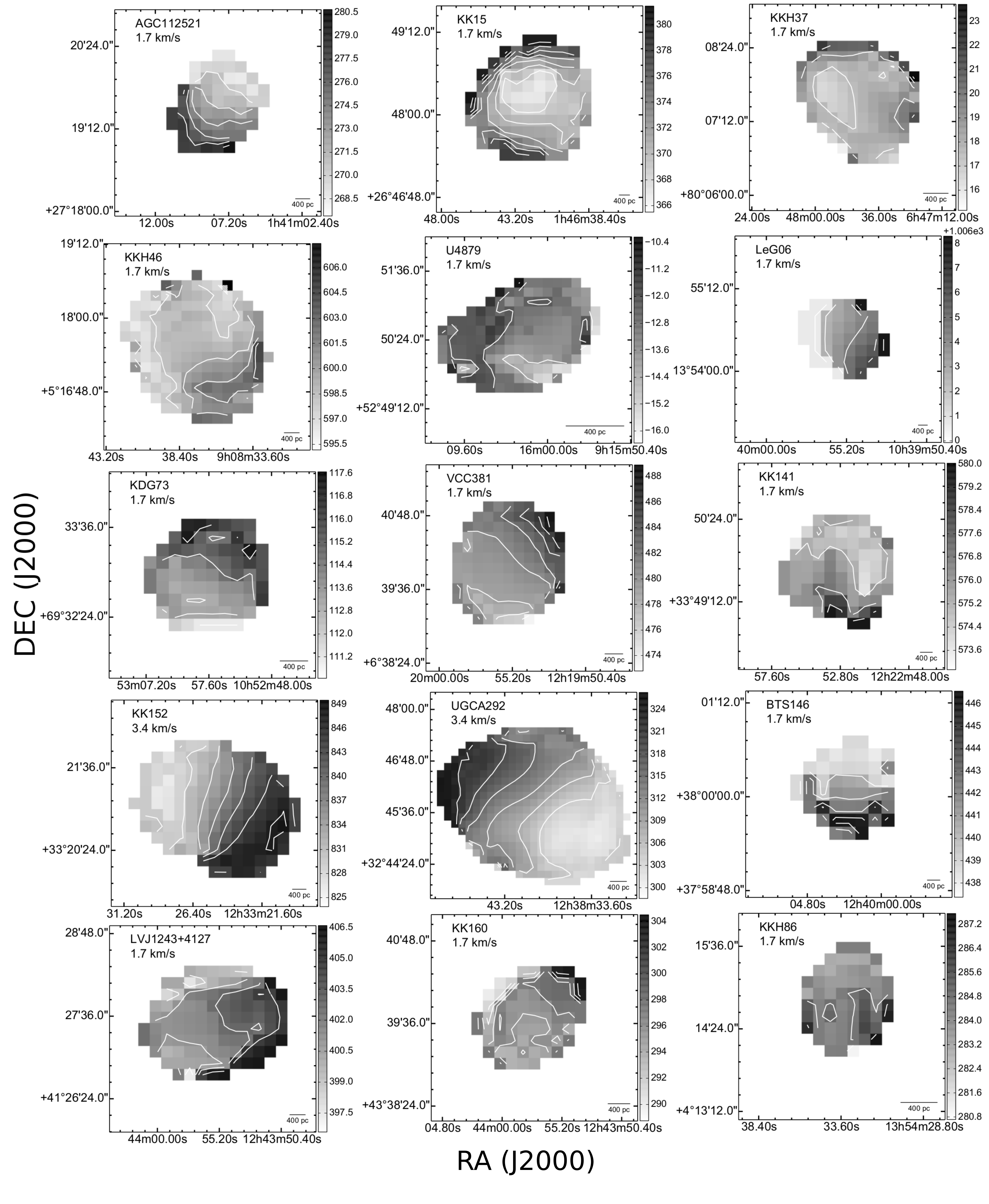}}
\end{tabular}
\end{center}
\caption{Velocity field of our detected galaxies. The spacing between subsequent contours were quoted at the upper left corner of every panel. Almost all our sample galaxies show ordered rotation in their velocity map.}
\label{momnt1}
\end{figure*}







\begin{table*}
\centering
\caption{Observation details}
\begin{tabular}{|l|c|c|c|c|c|c|c|c|c|}
\hline
Galaxy name & Date of observations & velocity coverage & Time on source & Synthesized beam & Single channel rms \\
& & (\kms) & (Hr) & (arcsec$^2$) & (mJy/beam)\\
(1) & (2) & (3) & (4) & (5) & (6) \\
\hline
AGC112521 & December 10, 2010 & $-145 - 734$ & $6$ & $40.64 \times 35.75, ~27.85 \times 22.56, ~13.91 \times 10.55$ & $2.0, ~1.5, ~1.3$\\ 
KK15	 & November 14, 2010 & $5 - 886$ & $4$ & $44.28 \times 36.50, ~27.35 \times 24.03, ~15.02 \times 11.39$ & $1.4, ~1.1, ~0.9$ \\
KKH37	 & December 29, 2010 & $-643 - 234$ & $5.3$ & $54.40 \times 35.55, ~25.98 \times 19.19, ~12.14 \times 9.85$ & $3.5, ~2.6, ~1.6$ \\
KKH46	 & December 10, 2010 & $251 - 1133$ & $3.8$ & $43.26 \times 36.75, ~30.03 \times 25.91, ~26.54 \times 10.51$ & $3.0, ~2.7, ~2.0$ \\
UGC4879 & August 06, 2010 & $-154 - 56$ & $3.8$ & $48.16 \times 34.03, ~34.18 \times 21.52, ~25.02 \times 14.78$ & $3.8, ~3.2, ~2.8$ \\
LeG06 	 & October 15, 2010 & $831 - 1272$ & $6.8$ & $45.00 \times 38.07, ~26.95 \times 22.50, ~12.40 \times 10.71$ & $3.8, ~3.0, ~1.1$ \\
KDG073	 & March 14, 2009 & $-19 - 191$ & $6.75$ & $45.32 \times 35.45, ~28.42 \times 22.00, ~14.44 \times 10.60$ & $2.8, ~1.7, ~1.5$ \\
VCC0381	 & August 08, 2010 & $273 - 714$ & $4.5$ & $45.28 \times 35.57, ~31.87 \times 23.50, ~23.40 \times 10.27$ & $3.1, ~2.9, ~2.5$ \\
KK141	 & November 14, 2010 & $37 - 919$ & $4.5$ & $44.49 \times 35.97, ~30.14 \times 24.39, ~13.31 \times 9.38$ & $2.1, ~1.8, ~1.5$ \\
KK152	 & August 09, 2010 & $494 - 1377$ & $4.5$ & $44.21 \times 33.60, ~29.53 \times 21.46, ~16.15 \times 9.71$ & $3.7, ~3.2, ~2.5$ \\
UGCA292	 & December 10, 2010 & $-171 - 708$ & $4.5$ & $45.22 \times 35.23, ~27.79 \times 23.95, ~15.09 \times 11.84$ & $2.6, ~2.5, ~1.9$\\ 
BTS146	 & December 11, 2010 & $39 - 920$ & $5.25$ & $44.28 \times 34.73, ~30.92 \times 21.58, ~16.29 \times 11.23$ & $1.1, ~0.8, ~0.7$ \\
LVJ1243+4127 & January 02, 2011 & $-69 - 811$ & $3.75$ & $49.86 \times 35.72, ~26.71 \times 20.43, ~13.92 \times 10.12$ & $3.2, ~2.6, ~2.0$ \\
KK160	 & December 31, 2010 & $-104 - 775$ & $4.4$ & $49.27 \times 35.55, ~28.43 \times 21.65, ~14.01 \times 9.93$ & $2.9, ~2.3, ~1.5$ \\
KKH86	 & November 13, 2008 & $181 - 392$ & $5.25$ & $43.20 \times 35.01, ~34.17 \times 23.83, ~29.53 \times 14.45$ & $2.6, ~2.3, ~1.8$ \\
LeG18	 & December 11, 2010 & $466 - 1350$ & $3.75$ & $87.52 \times 35.10, ~73.18 \times 23.85, ~62.30 \times 9.21$ & $7.4, ~9.0, ~6.3$ \\
PGC1424345 & August 12, 2010 & $623 - 1064$ & $4.5$ & $70.04 \times 33.61, ~59.18 \times 20.04, ~46.17 \times 16.54$ & $7.5, ~8.3, ~20.5$ \\
KDG090 & March 14, 2009 & $155 - 366$ & $3.3$ & $70.04 \times 33.61, ~59.18 \times 20.04, ~46.17 \times 16.54$ & $6.3, ~4.1, ~2.9$ \\
LVJ1217+4703	 & August 07, 2010 & $183 - 623$ & $4.4$ & $49.20 \times 37.41, ~47.16 \times 35.0, ~44.06 \times 33.61$ & $4.7, ~7.7, ~12.1$ \\
KK138 & December 31, 2010 & $39 - 920$ & $4.5$ & $42.10 \times 40.14, ~27.50 \times 23.83, ~11.91 \times 9.51$ & $1.8, ~1.7, ~1.7$ \\
KK191   & August 13, 2010 & $3 - 884$ & $4.5$ & $56.76 \times 34.50, ~46.54 \times 18.22, ~33.38 \times 11.72$ & $4.9, ~4.3, ~6.6$ \\
\hline
\end{tabular}
\label{table2_figgs}
\end{table*}

\section{Observation \& data analysis}

For all our observations we used the newly installed GMRT Software Back-end (GSB). A bandwidth of 2.08 MHz with 256 channels or a bandwidth of 4.17 MHz with 512 channels were used keeping the spectral resolution constant at $\sim$ 8.1 KHz (velocity width of $\sim$ 1.7 \kms). In every observing run flux calibration and bandpass calibration were done by observing standard flux calibrators 3C48, 3C147 or 3C286 at the starting and at the end of the observation. The phase calibration were done by observing a phase calibrator from the VLA list of calibrators within an angular distance of $\lesssim~10^o$ of the source once in every 45 minutes. 

Typically about 6 hrs of time was alloted for a single observation, with the actual on-source time varying between $\sim$ 2-5~hrs. All data were reduced in classic \scalebox{.7}{AIPS}. For every galaxy, phase and bandpass calibration was done after editing bad visibilities. Online doppler tracking was not done during observation, hence the data were corrected for earth's motion using \scalebox{.7}{AIPS} task \scalebox{.7}{CVEL}. The GMRT has a hybrid configuration \citep{swarup91} with 12 antennas inside the central square (2 km $\times$ 2 km) and 18 antennas spread over $\sim$ 25 km area in an approximate ``Y'' shaped array. Due to its hybrid configuration, GMRT is capable of sampling both the small and large angular scales within a single observing run. The longest achievable baseline at 21cm wavelength is $\sim$ 120 k$\lambda$. 

Dirty image cubes at different resolutions were made using the task \scalebox{.7}{IMAGR} in \scalebox{.7}{AIPS} by using {\it `Natural'} and {\it `Robust'} weighting schemes with different values of uvrange and uvtaper. While the {\it `Natural'} weighting maximizes the signal to noise ratio, it is know to produce non-gaussian beam profiles and induces complex noise properties into the image. Whereas, {\it `Robust'} weighting scheme produces somewhat better behaving beam profiles with a diminished SNR. As FIGGS2 sample galaxies are ultra-faint,
 and a high SNR map favours manual inspection/investigation, we show only {\it `Natural'} weighted maps in further analysis, though we produced maps using both the weighting schemes. The low resolution dirty cubes were inspected to identify the channels containing \HI emission. Since the emission is faint, we found it very difficult and subjective to generate masks for cleaning or generating moment maps. Prior to this we used the line-free channels (identified in the low resolution cube) to fit and subtract the continuum in the image plane using the task \scalebox{.7}{IMLIN} in \scalebox{.7}{AIPS}. The continuum subtracted cubes were then cleaned up to an rms level of $\sim 2.5$ times single channel rms (line free) using the task \scalebox{.7}{APCLN}. We also tried a multi-scale cleaning but this did not significantly improve the quality of images. Although all of our observations were carried out with a velocity resolution of $\sim 1.65$ \kms, we collapsed adjacent channels (reducing velocity resolution) to increase SNR wherever necessary.

Moment maps were made using the task \scalebox{.7}{MOMNT} in classic \scalebox{.7}{AIPS}. We smoothed the data using a Gaussian kernel of width 6 pixels in spatial coordinates and a Hanning smoothing of width 3 pixels were applied to the velocity coordinates. We apply a cut off of 1.5-2 times the per channel rms to select emission regions to be included in the moment maps. Total intensity images at different resolutions provide complementary information. For example the effect of local processes like star formation, feedback etc. are best studied using high resolution images, whereas the large-scale dynamics, global extent of \HI, dark matter halo properties etc. are better studied using low resolution images. As an example, in Figure~\ref{momres} we show integrated \HI emission images of one of the FIGGS2 sample galaxies, (viz. UGC~4879) at different spatial resolutions. The galaxy shows a faint extended structure at the south-east corner in low resolution image (panel [A]) which is resolved out at higher resolution. On the other hand, the fine details of  the morphology of the galaxy in the central region can be more clearly seen in the high resolution images. 

We detected \HI emission in 15 out of 20 galaxies. Two (LeG18, LVJ1217+4703) out of the five non-detections have quite large single-dish peak fluxes ($>$ 25 mJy \citep{huchtmeier09}). However their GMRT observations were affected by strong RFI and a significant fraction of the data had to be flagged, resulting in higher noise levels in the data cube. Despite the increased noise level, one would have expected to detect the \HI emissions at least at 3$\sigma$ level, and hence the non detections are surprising, if the single dish fluxes are correct. The reason for this discrepancy is unclear to us. Though the quoted single dish flux of KDG90 is quite high ($\sim~$23.6 Jy\kms \citep{koribalski04}), this dSph galaxy resides within $\sim 10^{\prime}$ of the bright spiral NGC4214 having \HI flux of 147 Jy\kms and Holmberg diameter of 8.5 arcmin. Hence, most likely this is a case of \HI confusion under single-dish observation. Subsequently, VLA observations (VLA-ANGST survey, \citep{ott12}) also did not detect any emission from this galaxy. The single dish \HI spectra for KK138 has a velocity width of 186 \kms\ and a very low peak flux of $\sim$10 mJy. Such a large velocity width is not expected for dwarf galaxy; it seems likely that the single dish detection is spurious. In the case of KK191 there is a large spiral galaxy NGC5055 within an angular distance of $\sim~25^{\prime}$. NGC5055 has a central velocity of 510 \kms\ and a velocity width of $\sim$ 400 \kms\ which overlaps with the quoted velocity for KK191, i.e. 368 \kms \citep{huchtmeier00}. Hence it is possible that the single dish detection is confused. The observation details and analysis results are presented in Table~\ref{table2_figgs}. The columns are as follows: column (1): the galaxy name, column (2): date of observation, column (3): the velocity (heliocentric) coverage of the observing band, column (4): on-source time in hour, column (5): synthesized beam size at different resolution data cubes, column (6): corresponding single channel rms. 

In Fig.~\ref{spec} we overplot the \HI global spectra extracted from our observations (red solid lines) on top of the single dish spectra (blue dashed line) (wherever available, see \S~\ref{resultsanddiscussion} for more details) of our detected galaxies. From the Figure, it can be seen that almost in all the cases, our observed spectra recovers less flux as compared to the single-dish flux. For example, the synthesis observation of UGC04879 using WSRT \citep{bellazzini11} recovers much more \HI flux (2.2$\pm$0.1 Jy\kms) than what is recovered by the GMRT (1.35$\pm$0.7 Jy\kms). We expect that this is because the GMRT has fewer short spacings than the WSRT and resolves out most of the low column density extended emission. We have carefully checked our calibration solutions and compared the recovered secondary calibrator fluxes with VLA calibrator manual. In all the cases our fluxes match the catalog value within 10\%. A 10\% error in calibration is insufficient to explain the flux discrepancies between GMRT spectra and the single-dish spectra.

\section{Results \& Discussion}
\label{resultsanddiscussion}
 In Figure~\ref{ovrplot} we show the integrated \HI distribution (contours) overlayed on the optical images for the detected galaxies. The lowest contour levels are quoted at the upper left corners of each panel in the unit of \acc. We used optical images from  SDSS survey (`g' band) whenever available or else we use images from DSS survey (`B' band). We quote the source of the optical images at the top right corner of each panel. To compare the \HI and optical extents and to show large scale \HI structures of our sample galaxies, we choose to overlay low resolution (higher SNR) \HI maps on top of the optical images in Fig.~\ref{ovrplot}. However, due to non-uniform sampling of the visibility plane across our sample, the synthesised beams vary considerably for galaxy to galaxy even after setting the same maximum range of visibility (5 kilo $\lambda$) during imaging. The synthesised beams are shown at the left bottom corner of every panel. We note that the optical center and the \HI center of many galaxies do not coincide (e.g U4879, KKH86, LVJ1243+4127). We speculate that feedback from star formation could be a possible cause of these offsets. 

In Figure~\ref{spec} we plot the \HI global spectra of our detected galaxies (red solid line). As the detected galaxies are very faint, the global spectra at $\sim$ 1.8 \kms\ resolution some times has a very low SNR. Hence adjacent channels were collapsed together to increase SNR wherever necessary. The velocity resolutions used for different galaxies are quoted at the upper left corner of the respective panels in Figure~\ref{spec}. We also over-plot the single-dish spectra (blue dashed line) for comparison. For KKH37 and UGC04879 we could not find single dish spectra from literature. For BTS146, we note that there is a difference in the central heliocentric velocity ($V_{sys}$) between single dish spectra and the GMRT spectra. However, \cite{kovac09} observed the same galaxy using WSRT and found a central velocity of 446 $\pm$ 17 \kms which matches well what we found ($\sim$ 440 \kms). 

The parameters derived from the global spectra are listed in Table \ref{table3_figgs}. The columns are as follows: column (1) the galaxy name, column (2) The integrated \HI flux, column (3) systematic velocity ($V_{sys}$), column (4) the velocity width at 50 percent of the peak flux ($\Delta_{50}$), column (5) The \HI diameter derived by ellipse fitting at a column density, $\rm N_{HI}= 0.3 \ M_{\odot} /pc^{2}$, column (6) the ratio of the \HI diameter to the optical diameter, column (7) the derived \HI mass, column (8) mass to light ratio ($M_{HI}/L_B$), column (9) the ratio of GMRT flux to single-dish flux, column (10) \HI inclination assuming an intrinsic thickness of 0.6 \citep{roychowdhury10}. The associated errors are quoted along with the derived parameters. The $V_{sys}$ and the $\Delta_{50}$ were derived by fitting a Gaussian profile to the global \HI spectra. The quoted errors on $V_{sys}$ and $\Delta_{50}$ represent fitting errors only. We estimate the \HI diameter by fitting an ellipse to the iso-\HI column density contour at $\rm N_{HI}= 0.3 \ M_{\odot} /pc^{2}$. The errors in the estimation of \HI diameter ($D_{HI}$) is expected to be dominated by the errors in the \HI map. To account this, we first compute an error map by using the knowledge of the rms in the \HI cube and the number of channels used to make the \HI map. We then estimate a typical error involved in measured column density at $\rm N_{HI}= 0.3 \ M_{\odot} /pc^{2}$ contours (i.e. the mean error along the $\rm N_{HI}= 0.3 \ M_{\odot} /pc^{2}$ contour from the error map). We then construct 1000 realization of \nh~which are consistent with $\rm N_{HI}= 0.3 \ M_{\odot} /pc^{2}$ within the error. We use these \nh~values for \HI isophotes and fit ellipses to these isophotes. We use the standard deviation as an estimate of the errors in the fit parameters. The errors in $\rm D_{HI}$, $\rm D_{HI}/D_{opt}$ and $\rm i_{HI}$ were estimated in this way.

\begin{table*}
\centering
\caption{Results from the GMRT observations of FIGGS2 sample galaxies}
\begin{tabular}{|l|c|c|c|c|c|c|c|c|c|}
\hline
Galaxy & $\rm FI_{GMRT}$ & $\rm V_{sys}$ & $\rm \Delta V_{50}$ & $\rm D_{HI}$ & $\rm D_{HI}/D_{opt}$ & $\rm M_{HI}$ & $\rm M_{HI}/L_B$ & $\rm FI_{GMRT}/FI_{SD}$ & $\rm i_{HI}$\\
 & (Jy \kms) & (\kms) & (\kms) & (arcmin) & & $(\times 10^{7} \ M_{\odot})$ & & & ($^o$) \\
 (1) & (2) & (3) & (4) & (5) & (6) & (7) & (8) & (9) & (10) \\
\hline
AGC112521 & $0.44 \pm 0.34$ & $270.4 \pm  0.2$ & $25.0 \pm  3.8$ & $1.12 \pm 0.14$ & $ 1.9 \pm  0.2$ & $0.38 \pm 0.29$ & $0.67 \pm 0.51$ & $ 0.7 \pm  0.5$ & $44 \pm 7$ \\
KK15 & $0.52 \pm 0.19$ & $371.3 \pm  1.3$ & $23.8 \pm  3.1$ & $0.93 \pm 0.14$ & $ 1.6 \pm  0.2$ & $0.92 \pm 0.34$ & $1.12 \pm 0.42$ & $ 0.6 \pm  0.2$ & $63 \pm 6$ \\
KKH37 & $0.70 \pm 0.13$ & $17.4 \pm  0.1$ & $17.2 \pm  0.9$ & $1.46 \pm 0.12$ & $ 1.3 \pm  0.1$ & $0.20 \pm 0.04$ & $0.29 \pm 0.05$ & $ 0.4 \pm  0.1$ & $64 \pm 4$ \\
KKH46 & $1.96 \pm 0.47$ & $598.2 \pm  0.3$ & $21.2 \pm  0.8$ & $1.88 \pm 0.23$ & $ 3.1 \pm  0.4$ & $2.07 \pm 0.49$ & $1.59 \pm 0.38$ & $ 0.8 \pm  0.2$ & $39 \pm 4$ \\
UGC04879 & $1.35 \pm 0.66$ & $-13.2 \pm  0.2$ & $14.2 \pm  1.2$ & $1.36 \pm 0.37$ & $ 0.4 \pm  0.1$ & $0.06 \pm 0.03$ & $0.06 \pm 0.03$ & $ 0.5 \pm  0.3$ & $46 \pm 7$ \\
LeG06 & $0.22 \pm 0.37$ & $1005.9 \pm  2.6$ & $16.3 \pm  6.6$ & $0.54 \pm 0.31$ & $ 0.9 \pm  0.5$ & $0.56 \pm 0.94$ & $0.62 \pm 1.04$ & $ 0.8 \pm  1.3$ & $54 \pm 17$ \\
KDG073 & $0.40 \pm 0.18$ & $114.6 \pm  0.5$ & $14.2 \pm  1.2$ & $1.23 \pm 0.30$ & $ 1.0 \pm  0.2$ & $0.13 \pm 0.06$ & $0.39 \pm 0.18$ & $ 0.4 \pm  0.2$ & $71 \pm 14$ \\
VCC0381 & $1.07 \pm 0.30$ & $479.8 \pm  0.2$ & $22.9 \pm  1.3$ & $1.45 \pm 0.11$ & $ 1.9 \pm  0.1$ & $0.56 \pm 0.16$ & $0.74 \pm 0.21$ & $ 0.4 \pm  0.1$ & $38 \pm 5$ \\
KK141 & $0.43 \pm 0.18$ & $576.0 \pm  0.8$ & $14.5 \pm  1.8$ & $0.98 \pm 0.19$ & $ 2.4 \pm  0.5$ & $0.61 \pm 0.26$ & $0.98 \pm 0.41$ & $ 0.4 \pm  0.2$ & $45 \pm 11$ \\
KK152 & $1.78 \pm 0.37$ & $834.7 \pm  0.9$ & $30.5 \pm  2.0$ & $1.63 \pm 0.18$ & $ 1.5 \pm  0.2$ & $2.00 \pm 0.42$ & $0.80 \pm 0.17$ & $ 0.6 \pm  0.1$ & $66 \pm 4$ \\
UGCA292 & $11.67 \pm 0.62$ & $309.2 \pm  0.1$ & $24.6 \pm  0.3$ & $3.12 \pm 0.22$ & $ 3.1 \pm  0.2$ & $4.08 \pm 0.22$ & $4.51 \pm 0.24$ & $ 1.3 \pm  0.1$ & $37 \pm 4$ \\
BTS146 & $0.39 \pm 0.15$ & $440.5 \pm  1.8$ & $25.5 \pm  4.3$ & $1.00 \pm 0.15$ & $ 2.9 \pm  0.4$ & $0.66 \pm 0.26$ & $0.56 \pm 0.22$ & $ 0.7 \pm  0.3$ & $59 \pm 7$ \\
LVJ1243+4127 & $0.62 \pm 0.53$ & $403.2 \pm  0.0$ & $16.5 \pm  2.6$ & $1.22 \pm 0.20$ & $ 0.9 \pm  0.1$ & $0.54 \pm 0.46$ & $0.66 \pm 0.56$ & $ 0.5 \pm  0.4$ & $68 \pm 5$ \\
KK160 & $0.51 \pm 0.53$ & $301.6 \pm  0.1$ & $20.0 \pm  3.4$ & $1.45 \pm 0.31$ & $ 2.5 \pm  0.5$ & $0.22 \pm 0.23$ & $0.62 \pm 0.65$ & $ 0.6 \pm  0.6$ & $71 \pm 8$ \\
KKH86 & $0.45 \pm 0.16$ & $285.0 \pm  0.7$ & $15.1 \pm  1.5$ & $1.11 \pm 0.22$ & $ 1.3 \pm  0.3$ & $0.07 \pm 0.03$ & $0.35 \pm 0.13$ & $ 0.9 \pm  0.3$ & $59 \pm 9$ \\
\hline
\end{tabular}
\label{table3_figgs}
\end{table*}

In Figure~\ref{momnt1} we present the velocity fields of the detected galaxies. We note that in many cases emission has been detected only across a few channels. As the SNR is poor, we did not take a Gaussian-Hermite polynomial fitting approach to derive the velocity field. Instead we adopted the intensity weighted first moment of the spectral cube as the velocity field. From Figure~\ref{momnt1}, we can see that, there are ordered velocity fields which is an indication of rotation in many galaxies (e.g. AGC112521, LeG06, KDG73, VCC381). But at the same time there are a few galaxies in the sample which show chaotic velocity fields, for example, KKH86, KK160, KKH37. The chaotic appearance of the velocity field could be due to the low SNR and low spatial resolution in the spectral cube. For the same reasons, the PV diagrams are noisy and do not bring out kinematics of the galaxies and hence we do not present them here.

\begin{figure}
\begin{center}
\resizebox{85mm}{!}{\includegraphics{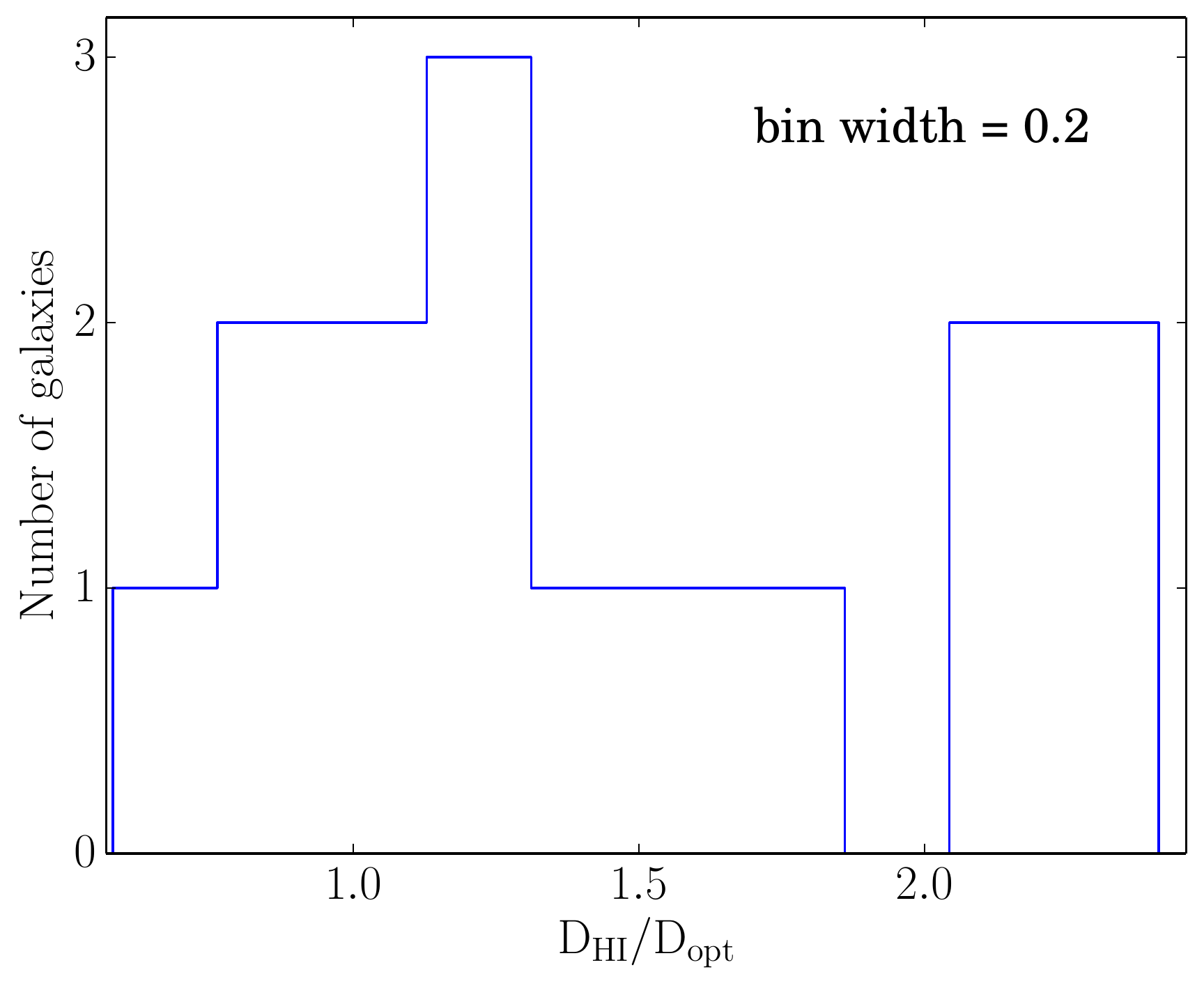}}
\end{center}
\caption{Histogram of \HI diameters of our sample galaxies normalized to optical diameter. One can see that almost all our galaxies have \HI diameter larger than the optical diameter except one (UGC4879). See the text for more discussion.}
\label{dh1_dopt}
\end{figure}

In Figure~\ref{dh1_dopt} we plot the histogram of \HI diameters of our sample galaxies. To compare the extent of \HI disks with their optical counterparts, we normalised the \HI diameter by the optical diameter ($\rm D_{opt}$) of the galaxies. Isophotal radii e.g. $R_{Holm}$ or $R_{25}$ have limited meaning for dwarf galaxies having low surface brightness. These radii estimates could be prone to systematic under-estimation of their optical extent. Hence we perform photometric analysis of B-band image of our galaxies, and fit the surface brightness profiles with an exponential profile. Adopting a convention by \citep{swaters02}, we define optical radii as 3.2 times exponential scale length. However for four of our detected galaxies (KKH37, LeG06, KDG073 and KKH86), optical photometry (in B band) could not be performed due to poor quality of available data. For these galaxies, we considered Holmberg radius as optical radii. In many previous \HI surveys \citep{broeils97,verheijen01,swaters02,noordermeer05} an isophote of $\rm 1 \ M_{\odot} pc^{-2}$ was adopted for ellipse fitting and estimating the \HI radii. However, most of our detected galaxies, fall short of \HI surface density of $\rm 1 \ M_{\odot} pc^{-2}$ even at the center. We have used an face-on \HI surface density of $\rm 0.3 \ M_{\odot} pc^{-2} \ (3.75 \times 10^{19} \ atoms \ cm^{-2})$ isophote to estimate the \HI diameter. The mean value of normalised \HI diameter is 1.54 which is somewhat lower than the value found for the FIGGS \citep{begum08c} sample which is 2.40. This may be in part to the very faint outer emission being resolved out. From our data, we found that for all our sample galaxies, \HI disk extends more than the optical disk except one. For the galaxy UGC4879, the \HI disk found to be smaller than its optical counterpart. From Figure~\ref{ovrplot} (5th image) we note that, a faint extended \HI emission is seen in the south-east corner, which may be indicative of diffuse emission not picked up in our observations. It is worth noting that for UGC04879 the GMRT observation picks up only about 50\% of the single dish flux.

\begin{figure}
\begin{center}
\resizebox{85mm}{!}{\includegraphics{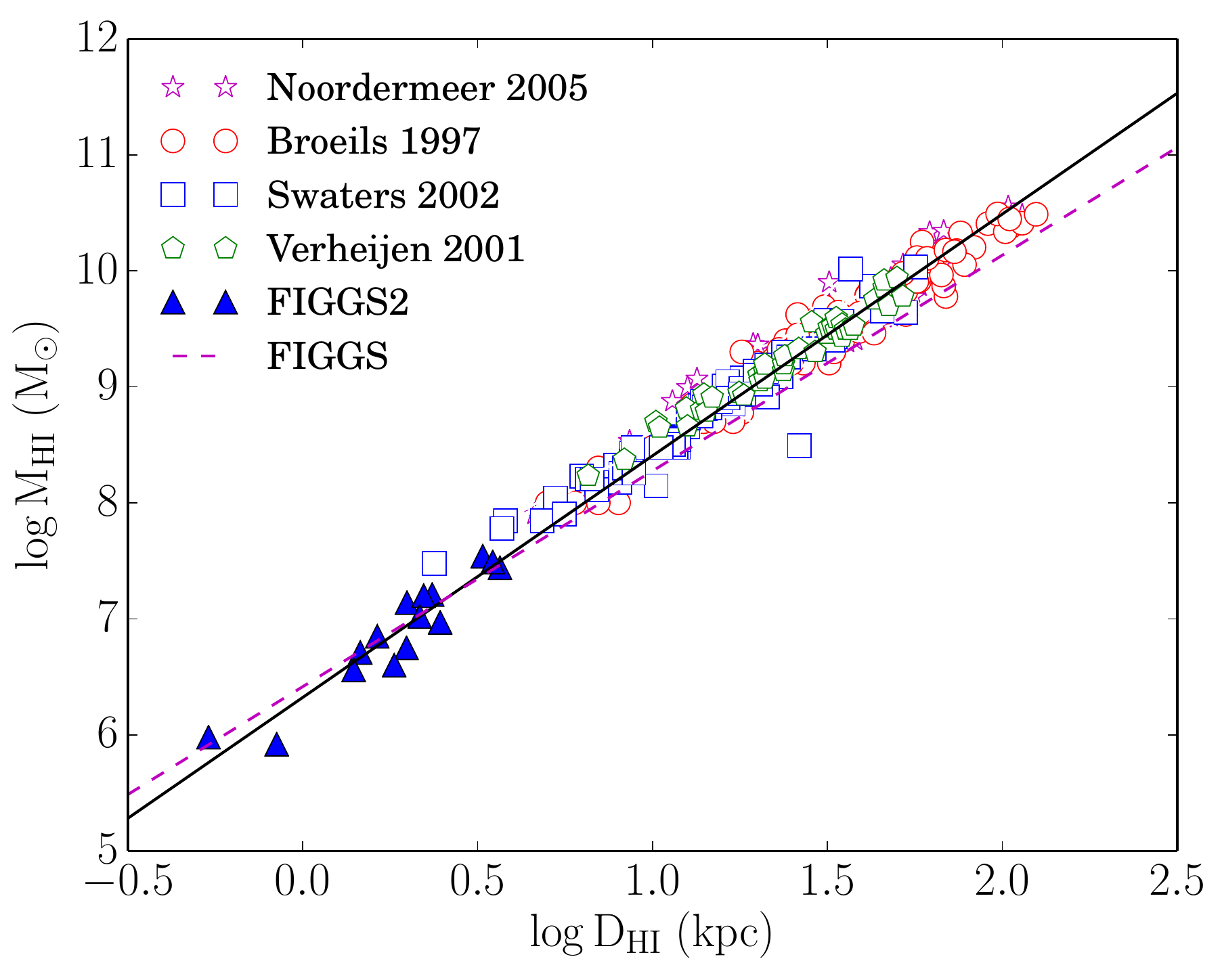}}
\end{center}
\caption{The \HI mass (single-dish) of the FIGGS2 sample as a function of \HI diameter (measured at a column density of $\rm 0.3 \ M_{\odot} pc^{-2} \ (3.75 \times 10^{19} \ atoms \ cm^{-2})$ ). The black solid line represents a straight line fit to the FIGGS2 data whereas the magenta dashed line represents a fit to the FIGGS data taken from \citep{begum08c}. The empty symbols in the plot represent data for spiral galaxies taken from literature. As the large spiral galaxies are bright in \HI the $\rm D_{HI}$ for them is defined at a column density of $\rm 1 \ M_{\odot} pc^{-2} \ (1.25 \times 10^{20} \ atoms \ cm^{-2})$. }
\label{mh1_dh1}
\end{figure}

The \HI diameter and the \HI mass of different types of galaxies exhibits a tight correlation. In Figure~\ref{mh1_dh1} we plot the correlation between the \HI diameter and the \HI mass of our sample galaxies (filled blue triangles). As the GMRT resolves out a significant amount of \HI at low column densities at the outer radii (as noted in \S4), we use single-dish \HI flux measurements in Fig.~\ref{mh1_dh1}. To compare the correlation with larger galaxies, we over plot data for spiral galaxies (\HI diameter defined at an \HI surface density of $\rm 1 \ M_{\odot} pc^{-2}$) from various previous \HI surveys \citep{broeils97,verheijen01,swaters02,noordermeer05}. The solid black line represents a linear fit to our (FIGGS2) data whereas the dashed magenta line represents a linear fit for FIGGS survey. It can be seen that due to the small size of our sample galaxies, our study extended this correlation to low mass and low diameter end. From the figure it can be noted that our data points follow the trend for spiral galaxies (hollow points) or for the FIGGS galaxies (magenta dashed line). However, we note that our data points might be affected by the facts that the $\rm D_{HI}$ were measured at a different \HI column density for FIGGS2 and for the spiral galaxies.

The best linear fit of $D_{HI}$ vs $M_{HI}$ correlation (black solid line) could be represented by

\begin{equation}
\log(M_{HI}) = (2.08 \pm 0.20) \log(D_{HI}) + (6.32 \pm 0.07)
\label{eq:mh_md}
\end{equation}

\noindent In Fig.\ref{mh1_dh1} the dashed magenta line represents the correlation for FIGGS galaxies. 
The slope and the intercept for FIGGS2 galaxies (i.e. $2.08 \pm 0.20$ and $6.32 \pm 0.07$) roughly matches with that of the FIGGS galaxies.

\begin{figure}
\begin{center}
\resizebox{85mm}{!}{\includegraphics{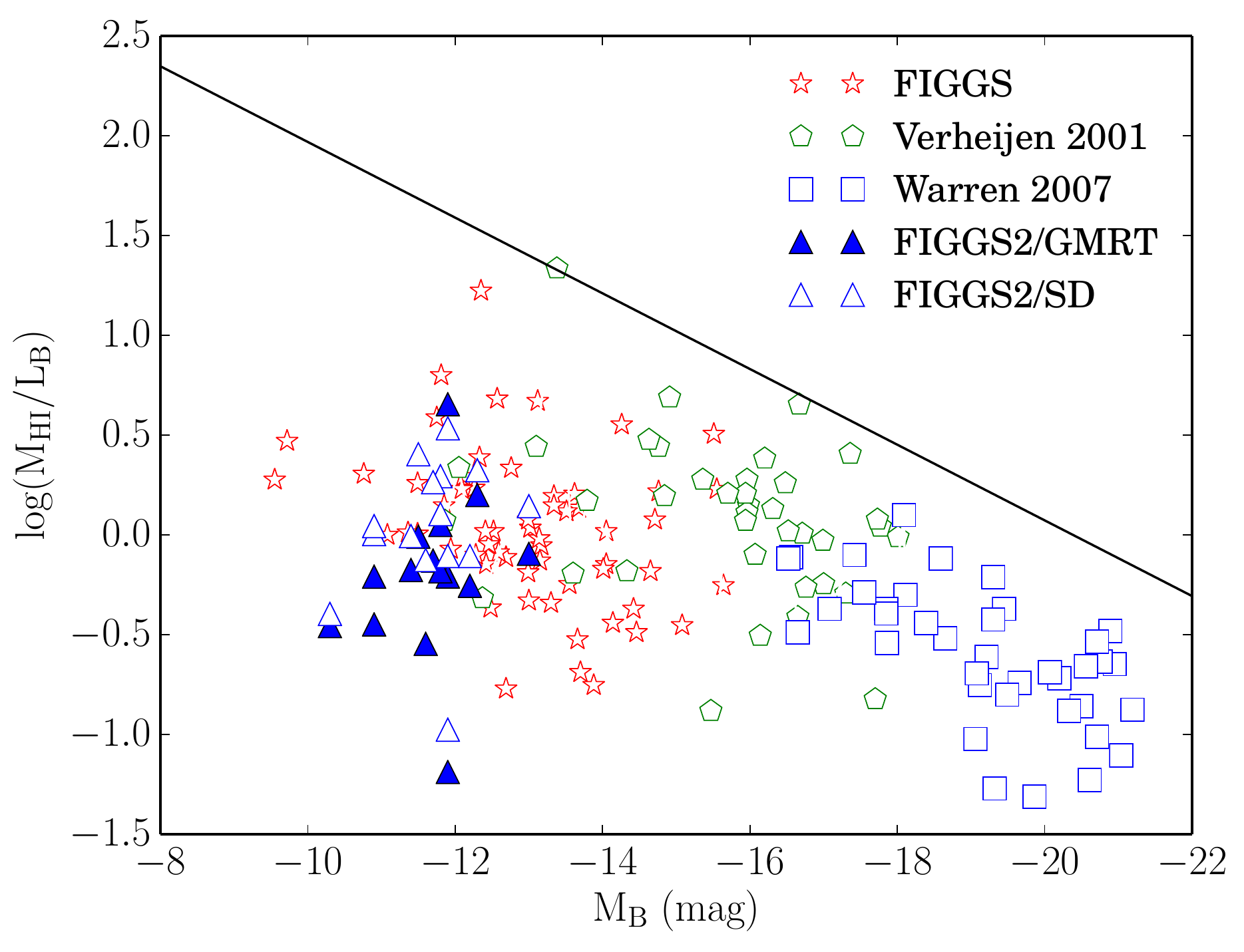}}
\end{center}
\caption{The log of \HI-mass-to-light ratio as a function of $\rm M_B$. Blue filled (GMRT \HI mass) and hollow (Single dish \HI mass) triangles are from FIGGS2, red hollow asterisks represent data from FIGGS survey whereas blue hollow squares and green hollow pentagons represent \citet{warren07} and \citet{verheijen01a} respectively. The solid line represents an empirically derived upper limit to $\rm M_{HI}/L_B$ from \citep{warren07}. See text for more details.}
\label{mh_mb}
\end{figure}

In Figure~\ref{mh_mb} we show the $\rm \log (M_{HI}/L_B)$ as a function of $\rm M_B$. Our sample galaxies are shown by filled (GMRT \HI mass) and hollow (Single dish \HI mass) blue triangles, whereas the red hollow asterisks represent the FIGGS sample. The blue hollow squares are from \citet{warren07} and green hollow pentagons are for galaxies from \citet{verheijen01a}. The solid line represents an empirically derived upper limit to the $\rm (M_{HI}/L_B)$ from \citet{warren07}. It can be thought of as a minimum fraction of the baryonic mass to be converted into stars in order to be stable under thermal equilibrium with gravity \citep{warren07} for a galaxy of given baryonic mass. It is interesting to note that all our sample galaxies lies well below the solid line (even with single-dish \HI mass). It implies that these small dwarf galaxies converted much more gas into stars than the minimum required to be stable under the balance of gravity and thermal energy.

In summary we have observed 20 faint galaxies with the GMRT to extend the FIGGS sample towards the low luminosity end. We  detected \hi emission from 15 of the galaxies. We find that these galaxies have the similar \hi\ mass to \hi\ diameter relation as the brighter dwarfs. These data will be useful for a host of studies of dwarf galaxies, including the interplay between gas and star formation, the phase structure of the atomic ISM, the structure and distribution of the dark matter halos, etc.

\section{Acknowledgements}
NNP would like to thank the anonymous referee for her/his valuable comments which helped to improve the paper significantly. NNP would like to thank the GMRT staff members for making the observations possible. The Giant Meter-wave Radio Telescope is run by the National Centre for Radio Astrophysics of the Tata Institute of Fundamental Research. IDK \& MES thanks for the support from Russian Foundation for Basic Research, grant 15-52-45004, and  the Russian Science Foundation, grant 14-02-00965.

\bibliographystyle{apalike}
\bibliography{bibliography}

\end{document}